# Axiomatic Tools versus Constructive approach to Unconventional Algorithms

Gordana Dodig-Crnkovic[1] and Mark Burgin[2]

**Abstract.** In this paper, we analyze axiomatic issues of unconventional computations from a methodological and philosophical point of view. We explain how the new models of algorithms changed the algorithmic universe, making it open and allowing increased flexibility and creativity. However, the greater power of new types of algorithms also brought the greater complexity of the algorithmic universe, demanding new tools for its study. That is why we analyze new powerful tools brought forth by the axiomatic theory of algorithms, automata and computation.

## 1 INTRODUCTION

Tradition in computation is represented by conventional computations. The conventional types and models of algorithms make the algorithmic universe, i.e., the world of all existing and possible algorithms, closed because there is a rigid boundary in this universe formed by recursive algorithms such as Turing machines.

Super-recursive algorithms controlling and directing unconventional computations break this boundary bringing people to an *open algorithmic universe* – a world of unbounded creativity. As the growth of possibilities involves much higher complexity of the new open world of super-recursive algorithms, innovative hardware and unconventional organization, we discuss means of navigation in this new open algorithmic world.

The paper is organized as follows. First in Section 2 we compare local and global mathematics. Section 3 addresses local logics and logical varieties, while Section 4 offers the discussion of projective mathematics versus reverse mathematics versus classical mathematics. Section 5 answers the question how to navigate in the algorithmic multiverse. Finally Section 6 presents our conclusions and provides directions for future work.

## 2 LOCAL MATHEMATICS VERSUS GLOBAL MATHEMATICS

Mathematics exists as an aggregate of various mathematical fields. If at the beginning, there were only two fields – arithmetic and geometry, now there are hundreds of mathematical fields and subfields. However, mathematicians always believed in mathematics as a unified system striving to build common and in some sense absolute foundations for all mathematical fields and subfields. At the end of the 19th century, mathematicians came very close to achieving this goal as the emerging set theory allowed building all mathematical structures using only sets and operations with sets. However, in the 20th century, it was discovered that there are different set theories. This brought some confusion and attempts to find the "true" set theory.

[1] School of Innovation, Design and Engineering, Mälardalen University, Sweden. Email: gordana.dodig-crnkovic@mdh.se
[2] Dept. of Mathematics, UCLA, Los Angeles, USA. Email: mburgin@math.ucla.edu

To overcome this confusion, Bell [1] introduced the concept of local mathematics in 1986. The fundamental idea was to abandon the unique absolute universe of sets central to the orthodox set-theoretic account of the foundations of mathematics, replacing it by a plurality of local mathematical frameworks. Bell suggested taking elementary toposes as such frameworks, which would serve as local replacements for the classical universe of sets. Having sufficient means for developing logic and mathematics, elementary toposes possess a sufficiently rich internal structure to enable a variety of mathematical concepts and assertions to be interpreted and manipulated. Mathematics interpreted in any such framework is called *local mathematics* and admissible transformation between frameworks amounts to a (definable) *change of local mathematics*. With the abandonment of the absolute universe of sets, mathematical concepts in general lose absolute meaning, while mathematical assertions liberate themselves from absolute truth values. Instead they possess such meanings or truth values only *locally*, i.e., *relative* to local frameworks. It means that the *reference* of any mathematical concept is accordingly not fixed, but *changes* with the choice of local mathematics.

It is possible to extend the approach of Bell in two directions. First, we can use an arbitrary category as a framework for developing mathematics. When an internal structure of such a framework is meager, the corresponding mathematics will be also indigent. Second, it is possible to take a theory of some structures instead of the classical universe of sets and develop mathematics in this framework.

A similar situation emerged in computer science.

Usually to study properties of computers and to develop more efficient applications, mathematicians and computer scientists use mathematical models. There is a variety of such models: Turing machines of different kinds (with one tape and one head, with several tapes, with several heads, with *n*-dimensional tapes, nondeterministic, probabilistic, and alternating Turing machines, Turing machines that take advice and Turing machines with oracle, etc.), Post productions, partial recursive functions, neural networks, finite automata of different kinds (automata without memory, autonomous automata, accepting automata, probabilistic automata, etc.), Minsky machines, normal Markov algorithms, Kolmogorov algorithms, formal grammars of different kinds (regular, context free, context sensitive, phrase-structure, etc.), Storage Modification Machines or simply, Shönhage machines, Random Access Machines (RAM), Petri nets, which like Turing machines have several forms – ordinary, regular, free, colored, self-modifying, etc.), and so on. All these models are constructive, i.e., they have a tractable explicit descriptions and simple rules for operation. Thus, the constructive approach is dominating in computer science.

This diversity of models is natural and useful because each of these classes is suited for some kind of problems. In other words, the diversity of problems that are solved by computers involves a corresponding diversity of models. For example, general problems of computability involve such models as Turing machines and partial recursive functions. Finite automata are used for text search, lexical analysis, and construction of semantics for programming languages. In addition, different computing devices demand corresponding mathematical models. For example, universal Turing machines and inductive Turing machines allows one to investigate characteristics of conventional computers [7]. Petri nets are useful for modeling and analysis of computer networks, distributed computation, and communication processes [31]. Finite automata model computer arithmetic. Neural networks reflect properties of the brain. Abstract vector and array machines model vector and array computers [7].

To utilize some kind of models that are related to a specific type of problems, we need to know their properties. In many cases, different classes have the same or similar properties. As a rule, such properties are proved for each class separately. Thus, alike proofs are repeated many times in similar situations involving various models and classes of algorithms.

In contrast to this, the *projective* (also called *multiglobal*) *axiomatic theory* of algorithms, automata and computation suggests a different approach [9][30]. Assuming some simple basic conditions (in the form of postulates, axioms and conditions), we derive in this theory many profound properties of algorithms. This allows one, when dealing with a specific model not to prove this property, but only to check the conditions from the assumption, which is much easier than to prove the property under consideration. In such a way, we can derive various characteristics of types of computers and software systems from the initial postulates, axioms and conditions.

Breaking the barrier of the Church-Turing Thesis drastically increased the variety of algorithmic model classes and changed the algorithmic universe of recursive algorithms to the multiverse of super-recursive algorithms, which consists of a plurality of local algorithmic universes. Each class of algorithmic models forms a local algorithmic universe, providing means for the development of local computer science in general and a local theory of algorithms in particular.

Local mathematics brings forth local logics because each local mathematical framework has its own logic and it is possible that different frameworks have different local logics.

## 3 LOCAL LOGICS AND LOGICAL VARIETIES

Barwise and Seligman (1997) developed a theory of information flow. In it, the concept of local logic plays a fundamental role in the modeling commonsense reasoning. The basic concept of this theory is a classification, which can be interpreted as a representation of some domain in the physical or abstract world. Each local logic corresponds to a definite classification. This implies a natural condition that each domain has its own local logic and different domains may have different local logics.

In the multiverse of super-recursive algorithms, each class of super-recursive algorithms forms a local algorithmic universe, which has a corresponding local logic. These logics may be essentially different. For instant, taking two local algorithmic universes formed by such classes as the class ***T*** of all Turing machines and the class ***TT*** of all total, i.e., everywhere defined, Turing machines, we can find that the first class satisfies the axiom of universality, which affirms existence of a universal algorithm, i.e., a universal Turing machine in this class. However, the class ***TT*** does not satisfy this axiom [9].

Analyzing the system of local logics, it is possible to see that there are different relations between them and it would be useful to combine these logics in a common structure. As it is explained in [9], local logics form a deductive logical variety or a deductive logical prevariety, which were introduced and studied in [4] as a tool to work with inconsistent systems of knowledge.

Minsky [24] was one of the first researchers in AI who attracted attention to the problem of inconsistent knowledge. He wrote that consistency is a delicate concept that assumes the absence of contradictions in systems of axioms. Minsky also suggested that in artificial intelligence (AI) systems this assumption was superfluous because there were no completely consistent AI systems. In his opinion, it is important to understand how people solve paradoxes, find a way out of a critical situation, learn from their own or others' mistakes or how they recognize and exclude different inconsistencies. In addition, Minsky [25] suggested that consistency and effectiveness may well be incompatible. He also writes [26]: "An entire generation of logical philosophers has thus wrongly tried to force their theories of mind to fit the rigid frames of formal logic. In doing that, they cut themselves off from the powerful new discoveries of computer science. Yes, it is true that we can describe the operation of a computer's hardware in terms of simple logical expressions. But no, we cannot use the same expressions to describe the meanings of that computer's output -- because that would require us to formalize those descriptions inside the same logical system. And this, I claim, is something we cannot do without violating that assumption of consistency." Then Minsky [26] continues, "In summary, there is no basis for assuming that humans are consistent - not is there any basic obstacle to making machines use inconsistent forms of reasoning". Moreover, it has been discovered that not only human knowledge but also representations/models of human knowledge (e.g., large knowledge bases) are inherently inconsistent [11]. Logical varieties or prevarieties provide powerful tools for working with inconsistent knowledge.

There are different types and kinds of logical varieties and prevarieties: *deductive* or *syntactic varieties* and *prevarieties*, *functional* or *semantic varieties* and *prevarieties* and *model* or *pragmatic varieties* and *prevarieties*. Syntactic varieties, prevarieties, and quasi-varieties (introduced in [10]) are built from logical calculi as buildings are built from blocks.

Let us consider a logical language $L$, an inference language $R$, a class **K** of syntactic logical calculi, a set $Q$ of inference rules ($Q \subseteq R$), and a class **F** of partial mappings from $L$ to $L$.

A triad **M** = ($A$, $H$, $M$), where $A$ and $M$ are sets of expressions that belong to $L$ ($A$ consists of axioms and $M$ consists of theorems) and $H$ is a set of inference rules, which belong to the set $R$, is called:

(1) a *projective syntactic* (**K**,**F**)-*prevariety* if there exists a set of logical calculi $C_i = (A_i, H_i, T_i)$ from **K** and a system of mappings $f_i : A_i \rightarrow L$ and $g_i : M_i \rightarrow L$ ($i \in I$) from **F** in which $A_i$ consists of all axioms and $M_i$ consists of all theorems of the logical calculus $C_i$, and for which the equalities $A = \cup_{i \in I} f_i(A_i)$, $H$

$= \cup_{i \in I} H_i$ and $M = \cup_{i \in I} g_i(M_i)$ are valid (it is possible that $C_i = C_j$ for some $i \neq j$).

(2) a *projective syntactic* (**K,F**)-*variety* with the depth $k$ if it is a projective syntactic (**K,F**)-quasi-prevariety and for any $i_1$, $i_2$, $i_3$, ... , $i_k \in I$ either the intersections $\cap_{j=1}^{k} f_{ij}(A_{ij})$ and $\cap_{j=1}^{k} g_{ij}(T_{ij})$ are empty or there exists a calculus $C = (A, H, T)$ from **K** and projections $f: A \rightarrow \cap_{j=1}^{k} f_{ij}(A_{ij})$ and $g: N \rightarrow \cap_{j=1}^{k} g_{ij}(M_{ij})$ from **F** where $N \subseteq T$;

(3) a *syntactic* **K**-prevariety if it is a projective syntactic (**K,F**)-prevariety in which $M_i = T_i$ for all $i \in I$ and all mappings $f_i$ and $g_i$ that define **M** are bijections on the sets $A_i$ and $M_i$, correspondingly;

(4) a *syntactic* **K**-*variety* if it is a projective syntactic (**K,F**)-variety in which $M_i = T_i$ for all $i \in I$ and all mappings $f_i$ and $g_i$ that define **M** are bijections on the sets $A_i$ and $M_i$, correspondingly.

The calculi $C_i$ used in the formation of the prevariety (variety) **M** are called *components* of **M**.

We see that the collection of mappings $f_i$ and $g_i$ makes a unified system called a prevariety or quasi-prevariety out of separate logical calculi $C_i$, while the collection of the intersections $\cap_{j=1}^{k} f_{ij}(A_{ij})$ and $\cap_{j=1}^{k} g_{ij}(T_{ij})$ makes a unified system called a variety out of separate logical calculi $C_i$. For instance, mappings $f_i$ and $g_i$ allow one to establish a correspondence between norms/laws that were used in one country during different periods of time or between norms/laws used in different countries.

The main goal of syntactic logical varieties is in presenting sets of formulas as a structured logical system using logical calculi, which have means for inference and other logical operations. Semantically, it allows one to describe a domain of interest, e.g., a database, knowledge of an individual or the text of a novel, by a syntactic logical variety dividing the domain in parts that allow representation by calculi.

In comparison with varieties and prevarieties, logical quasi-varieties and quasi-prevarieties studied in [5] are not necessarily closed under logical inference. This trait allows better flexibility in knowledge representation.

While syntactic logical varieties and prevarietis synthesize local logics in a unified system, semantic logical varieties and prevarieties studied in [5] unify local mathematics forming a holistic realm of mathematical knowledge.

In addition, syntactic logical varieties and prevarieties found diverse applications to databases and network technology (cf., for example, [6]).

## 4 PROJECTIVE MATHEMATICS VERSUS REVERSE MATHEMATICS VERSUS CLASSICAL MATHEMATICS

Mathematics suggests an approach for knowledge unification, namely, it is necessary to find axioms that characterize all theories in a specific area and to develop the theory in an axiomatic context. This approach worked well in a variety of mathematical fields.

Axiomatization has been often used in physics (Hilbert's sixth problem refers to axiomatization of branches of physics in which mathematics is prevalent), biology (The most enthusiastic proponent of this approach, the British biologist and logician Joseph Woodger, attempted to formalize the principles of biology—to derive them by deduction from a limited number of basic axioms and primitive terms—using the logical apparatus of the Principia Mathematica by Whitehead and Bertrand Russell, according to Britannica), and some other areas, such as philosophy or technology. It is interesting that the axiomatic approach was also used in areas that are very far from mathematics. For instance, Spinoza used this approach in philosophy, developing his ethical theories and writing his book Ethics in the axiomatic form. More recently, Kunii [20] developed an axiomatic system for cyberworlds.

With the advent of computers, deductive reasoning and axiomatic exposition have been delegated to computers, which performed theorem-proving, while the axiomatic approach has come to software technology and computer science. logical tools and axiomatic description has been used in computer science for different purposes. For instance, Manna [21] built an axiomatic theory of programs, while Milner [23] developed an axiomatic theory of communicating processes. An axiomatic description of programming languages was constructed by Meyer and Halpern [22]. Many researchers have developed different kinds of axiomatic recursion theories (cf., for example [15,19,14,13,29,28]).

However, in classical mathematics, axiomatization has the global character. Mathematicians tried to build a unique axiomatics for the foundations of mathematics. Logicians working in the theory of algorithms tried to find axioms comprising all models of algorithms.

This is the classical approach – axiomatizing the studied domain and then to deduce theorems from axioms. All classical mathematics is based on deduction as a method of logical reasoning and inference. *Deduction* is a type of reasoning processes that construct and/or evaluate *deductive arguments* and where the conclusion follows from the premises with logical necessity. In logic, an argument is called deductive when the truth of the conclusion is purported to follow necessarily or be a logical consequence of the assumptions. Deductive arguments are said to be valid or invalid, but never true or false. A deductive argument is valid if and only if the truth of the conclusion actually does follow necessarily from the assumptions. A valid deductive argument with true assumptions is called sound; a deductive argument which is invalid or has one or more false assumptions or both is called unsound. Thus, we may call classical mathematics by the name *deductive mathematics*.

The goal of deductive mathematics is to deduce theorems from axioms. Deduction of a theorem is also called proving the theorem. When mathematicians cannot prove some interesting and/or important conjecture, creative explorers invent new structures and methods, introducing new axioms to solve the problem. Researchers with a standard thinking try to prove that the problem is unsolvable.

Some consider deductive mathematics as a part of axiomatic mathematics, assuming that deduction (in a strict sense) is possible only in an axiomatic system. Others treat axiomatic mathematics as a part of deductive mathematics, assuming that there are other inference rules besides deduction.

While deductive mathematics is present in and actually dominates all fields of contemporary mathematics, reverse mathematics is the branch of mathematical logic that seeks to determine what are the minimal axioms (formalized conditions) needed to prove the particular theorem [17,18]. This direction in mathematical logic was founded by [15,16]. The method can briefly be described as going backwards from theorems to the axioms necessary to prove these theorems in some logical system [27]. It turns out that over a weak base theory, many mathematical statements are equivalent to the particular additional axiom needed to prove them. This methodology contrasts with the ordinary mathematical practice where theorems are deduced from a priori assumed axioms.

Reverse mathematics was prefigured by some results in set theory, such as the classical theorem that states that the axiom of choice, well-ordering principle of Zermelo, maximal chain priciple of Hausdorff, and statements of the vector basis theorem, Tychonov product theorem, and Zorn's lemma are equivalent over ZF set theory. The goal of reverse mathematics, however, is to study ordinary theorems of mathematics rather than possible axioms for set theory. A sufficiently weak base theory is adopted (usually, it is a subsystem of second-order arithmetic) and the search is for minimal additional axioms needed to prove some interesting/important mathematical statements. It has been found that in many cases these minimal additional axioms are equivalent to the particular statements they are used to prove.

Projective mathematics is a branch of mathematics similar to reverse mathematics, which aims to determine what are simple conditions needed to prove the particular theorem or to develop a particular theory. However, there are essential differences between these two directions: reverse mathematics is aimed at a logical analysis of mathematical statements, while projective mathematics is directed to making the scope of theoretical statements in general and mathematical statements in particular much larger and extending their applications. As a result, instead of proving similar results in various situations, it becomes possible to prove a corresponding general result in the axiomatic setting and to ascertain validity of this result for a particular case by demonstrating that all axioms (conditions) used in the proof are true for this case. In such a way the general result is projected on different situations. This direction in mathematics was founded by Burgin [9]. This approach contrasts with the conventional (deductive) mathematics where axioms describe some area or type of mathematical structures, while theorems are deduced from a priori assumed axioms.

Projective mathematics has its precursor in such results as extension of many theorems initially proved for numerical functions to functions in metric spaces or generalizations of properties of number systems to properties of groups, rings and other algebraic structures.

Here we use projective mathematics to study algorithms and automata. Our goal is to find some simple properties of algorithms and automata in general, to present these properties in a form of axioms, and to deduce from these axioms theorems that describe much more profound and sophisticated properties of algorithms. This allows one, taking some class **A** of algorithms, not to prove these theorems but only to check if the initial axioms are valid in **A**. If this is the case, then it makes possible to conclude that all corresponding theorems are true for the class **A**. As we know, computer scientists and mathematicians study and utilize a huge variety of different classes and types of algorithms, automata, and abstract machines. Consequently, such an axiomatic approach allows them to obtain many properties of studied algorithms and automata in a simple and easy way.

It is possible to explain goals of classical (deductive) mathematics, reverse mathematics and projective mathematics by means of relations between axioms and theorems.

A set *A* of axioms can be:

1. *Consistent* with some result (theorem) *T*, i.e., when the theorem *T* is added as a new axiom, the new system remains consistent, allowing in some cases to deduce (prove) this theorem.

2. *Sufficient* for some result (theorem) *T*, i.e., it is possible to deduce (prove) the theorem *T* using axioms from *A*.

3. *Irreducible* with respect to some result (theorem) *T*, i.e., the system *A* is a minimal set of axiom that allows one to deduce (prove) the theorem *T*.

After the discovery of non-Euclidean geometries, creation of modern algebra and construction of set theory, classical mathematics main interest has been in finding whether a statement *T* has been consistent with a given axiomatic system *A* (the logical goal) and then in proving this statement in the context of *A*. Thus, classical mathematics is concerned with the first relation. Reverse mathematics, as we can see, deals with the third relation.

In contrast to this, projective mathematics is oriented at the second relation. The goal is to find some simple properties of algorithms or automata in general, to present these properties in a form of a system *U* of axioms, and from these axioms, to deduce theorems that describe much more profound properties of algorithms and automata. This allows one, taking some class **A** of algorithms or automata, not to prove these theorems but only to check if all axioms from the system *U* are valid in **A**. If this is the case, then it is possible to conclude that all corresponding theorems are true for the class **A**. As we know, computer scientists and mathematicians study and utilize a huge variety of different classes and types of algorithms, automata, and abstract machines. Consequently, the projective axiomatic approach allows them to obtain many properties of studied algorithms in a simple and easy way. In such a way, the axiom system *U* provides a definite perspective on different classes and types of algorithms, automata, and abstract machines.

It is interesting that Bernays had a similar intuition with respect to axioms in mathematics, regarding them not as a system of statements about a subject matter but as a system of conditions for what might be called a relational structure. He wrote [2]:

"A main feature of Hilbert's axiomatization of geometry is that the axiomatic method is presented and practiced in the spirit of the abstract conception of mathematics that arose at the end of the nineteenth century and which has generally been adopted in modern mathematics. It consists in abstracting from the intuitive meaning of the terms . . . and in understanding the assertions (theorems) of the axiomatized theory in a hypothetical sense, that is, as holding true for any interpretation . . . for which the axioms are satisfied. Thus, an axiom system is regarded not as a system of statements about a subject matter but as a system of conditions for what might be called a relational structure . . . [On] this conception of axiomatics, . . . logical reasoning on the basis of the axioms is used not merely as a means of assisting intuition in the study of spatial figures; rather, logical dependencies are considered for their own sake, and it is insisted that in reasoning we should rely only on those properties of a figure that either are explicitly assumed or follow logically from the assumptions and axioms."

It is possible to formalize the approach of projective mathematics using logical varieties. Indeed, let us take a collection $C$ of postulates, axioms and conditions, which are formalized in a logical language as axioms. This allows us to assume that we have a logical variety $M$ that represents a given domain $D$ in a formal mathematical setting and contains the set $C$. For instance, the domain $D$ consists of a system of algorithmic models so that the logic of each model $D_i$ is a component $M_i$ of $M$. Then we deduce a theorem $T$ from the statements from $C$. Then instead of proving the theorem $T$ for each domain $D_i$, we check whether $C \subseteq M_i$. When this is true, we conclude that the theorem $T$ belongs to the component $M_i$ because $M_i$ is a calculus and thus, the theorem $T$ is valid for the model $D_i$. Because $C$ usually consists of simple statements, to check the inclusion $C \subseteq M_i$ is simpler than to prove $T$ in $M_i$.

## 5 HOW TO NAVIGATE IN THE ALGORITHMIC MULTIVERSE

It is possible to see that for a conformist, it is much easier to live in the closed algorithmic universe because all possible and impossible actions, as well as all solvable and insolvable problems can be measured against one of the most powerful and universal in the algorithmic universe classes of algorithms. Usually it has been done utilizing Turing machines.

Open world provides much more opportunities for actions and problem solving, but at the same time, it demands more work, more efforts and even more imagination for solving problems insolvable in the closed algorithmic universe. Even the closed algorithmic universe contains many classes and types of algorithms, which have been studied with a reference to a universal class of recursive algorithms. In some cases, partial recursive functions have been used. In other cases, unrestricted grammars have been employed. The most popular have been utilization of Turing machines. A big diversity of new and old classes of algorithms exist that demand specific tools for exploration.

Mathematics has invented such tools and one of the most efficient for dealing with diversity is the axiomatic method. This method has been also applied to the theory of algorithms, automata and computation when the axiomatic theory of algorithms, automata and computation was created [9]. In it, many profound properties of algorithms are derived based on some simple basic conditions (in the form of postulates, axioms and conditions). Namely, instead of proving similar results in various situations, it becomes possible to prove a necessary general result in the axiomatic setting and then to ascertain validity of this result for a particular case by demonstrating that all axioms (conditions) used in the proof are true for this case. In such a way the general result is projected on different situations. For instance, the theorem on undecidability of the Fixed Output Problem proved in [9] has more than 30 corollaries for various classes of algorithms, including the famous theorem about undecidability of the halting problem for Turing machines. Another theorem on recognizability of the Fixed Output Problem proved in [9] has more than 20 corollaries for various classes of algorithms, such as Turing machines, random access machines, Kolmogorov algorithms, Minsky machines, partial recursive functions, inductive Turing machines of the first order, periodic evolutionary Turing machines and limiting partial recursive functions.

The axiomatic context allows a researcher to explore not only individual algorithms and separate classes of algorithms and automata but also classes of classes of algorithms, automata, and computational processes. As a result, axiomatic approach goes higher in the hierarchy of computer and network models, reducing in such a way complexity of their study. The suggested axiomatic methodology is applied to evaluation of possibilities of computers, their software and their networks with the main emphasis on such properties as computability, decidability, and acceptability. In such a way, it became possible to derive various characteristics of types of computers and software systems from the initial postulates, axioms and conditions.

It is also worth mentioning that the axiomatic approach allowed researchers to prove the Church-Turing Thesis for an algorithmic class that satisfies very simple initial axioms [3,12]. These axioms form a system $C$ considered in the previous section and this system provides a definite perspective on different classes of algorithms, ensuring that in these classes the Church-Turing Thesis is true, i.e., it is a theorem.

Moreover, the axiomatic approach is efficient in exploring features of innovative hardware and unconventional organization.

It is interesting to remark that algorithms are used in mathematics and beyond as constructive tools of cognition. Algorithms are often opposed to non-constructive, e.g., descriptive, methods used in mathematics. Axiomatic approach is essentially descriptive because axioms describe properties of the studied objects in a formalized way.

Constructive mathematics is distinguished from its traditional counterpart, axiomatic classical mathematics, by the strict interpretation of the expression "there exists" (called in logic the *existential quantifier*) as "we can construct" and show the way how to do this. Assertions of existence should be backed up by constructions, and the properties of mathematical objects should be decidable in finitely many steps.

However, in some situations, descriptive methods can be more efficient than constructive tools. That is why descriptive methods in the form of the axiomatic approach came back to the theory of algorithms and computation, becoming efficient tool in computer science.

## 6 CONCLUSIONS

This paper demonstrated the role of the axiomatic methods for different paradigms of mathematics.

Classical mathematics utilizes global axiomatization and classical logic.

Local mathematics utilizes local axiomatization, diverse logics and logical varieties.

Reverse mathematics utilizes axiomatic properties decomposition and backward inference.

Projective mathematics utilizes view axiomatization, logical varieties and properties proliferation.

Here we considered only some consequences of new trends in the axiomatic approach to human cognition in general and mathematical cognition in particular. It would be interesting to study other consequences.

An important direction for future work is to study hardware systems and information processing architectures by applying the axiomatic methods of the mathematical theory of information technology [8].